\documentclass{ws-join}

\usepackage{placeins} 

\usepackage{xcolor}
\usepackage{xparse}

\usepackage{epsfig}

\usepackage{ulem} 
\usepackage{url} 
\begin{document}



\title{What is the potential for a second peak in the evolution of SARS-CoV-2 in emerging and developing economies? Insights from a SIRASD model considering the  informal economy
}


\author{Marcelo A. Pires$^1$\footnote{Corresponding author: piresma@cbpf.br}, Nuno Crokidakis$^2$, Daniel O. Cajueiro$^{3,4,5}$, Marcio Argollo de Menezes$^{2,3}$, Silvio M. Duarte~Queir\'{o}s$^{1,3}$ }

\address{$^1$Centro Brasileiro de Pesquisas F\'isicas, Rio de Janeiro/RJ, Brazil \\
$^2$Instituto de F\'isica, Universidade Federal Fluminense, Niter\'oi/RJ, Brazil \\
$^3$Instituto Nacional de Ci\^encia e Tecnologia de Sistemas Complexos INCT-SC, Rio de Janeiro/RJ, Brazil \\
$^4$Departamento de Economia, Universidade de Bras\'ilia, Bras\'ilia/DF, Brazil \\
$^5$Machine Learning Laboratory for Finance and Organizations, Universidade de Bras\'{i}lia/DF, Brazil
}

\maketitle


\begin{abstract}
We study the potential scenarios from a  Susceptible-Infected-Recovered-Asymptomatic-Symptomatic-Dead (SIRASD) model. As a novelty, we consider populations that differ in their degree of compliance with social distancing policies following socioeconomic attributes that are observed in emerging and developing countries. Considering epidemiological parameters estimated from  data of the propagation of  SARS-CoV-2 in Brazil -- where there is a significant stake of the population making their living in the informal economy and thus prone to not follow self-isolation -- we assert that if the confinement measures are lifted too soon, namely as much as one week of consecutive declining numbers of new cases, it is very likely the appearance of a second peak.  
Our approach should be valid for any country where the number of people involved in the informal economy is a large proportion of the total labor force. In summary, our results point out the crucial relevance of target policies for supporting people in the informal economy to properly comply  with preventive measures during the pandemic.

\end{abstract} 


\section{Introduction}

Despite 
existing some differences among the countries public health policies, the vast majority of them have tried to reduce the growth rate of the COVID-19 pandemic by implementing policies of social distancing\cite{Adam2020} aiming at preventing mayhem of the health-care systems, the so-called ``flattening of the curve''. 
A series of models have been brought forth to the specific study of the evolution of COVID-19 through the world 
 Initially, some of those works focused on its calibration in order to estimate typical parameters of the disease, like infection rates, epidemic doubling times among others\cite{crokidakis2020data,li2020substantial,muniz2020epidemic,liu2020reproductive,zhao2020preliminary,lai2020early,zhou2020preliminary,pedersen2020simple,tsallis2020predicting,rocha2020expected,weber2020trend}. After these preliminary studies, many authors considered the effect of
several types of non-pharmaceutical  interventions\cite{CrokidakisRJ,bastos2020modeling,de2020coronavirus,pellis2020challenges,manchein2020strong,maier2020effective,vasconcelos2020modelling,faggian2020proximity,ferguson2020impact,kraemer2020effect,bin2020fast,biswas2020covid,arenas2020mathematical}.

Despite the concerns related to public health, there are other impacts due to the implementation of social isolation policies. For example, it has been reported a decrease of $4.2\%$ in global $CO_{2}$ emission in the first quarter of 2020\cite{liu2020decreases}. In addition, we also observe economic impacts due to social distancing policies. Indeed, in some countries a considerable amount of the population is occupied with informal employment. These individuals, as well as people working in fundamental activities (hospitals, supermarkets, drugstores, and others) usually are not obeying the social distancing policies due to their professional activities.

A study analyzed the impacts of mobility lockdown in Italy due to the fast spreading of COVID-19\cite{bonaccorsi2020evidence} in which the authors identified two ways through which mobility restrictions affect the population. They verified that the impact of lockdown is stronger in municipalities with higher fiscal capacity, and also that mobility restrictions are stronger in municipalities for which inequality is higher and where individuals have lower income per capita, causing a segregation effect. In the same work~\cite{bonaccorsi2020evidence}, the authors also discussed about the income distribution, that plays an important role: municipalities where inequality is greater have experienced a stronger increase in mobility and their citizens are more at risk. Finally, they concluded that the results suggest the necessity of asymmetric fiscal measures. In other words, according to that work, central governments should implement financial transfer mechanisms to people, companies and local government in the form of living allowances, no-interest loans and treasury transfers to compensate the loss of tax income to allow each case to cope with the current scenario. As also stated therein, the absence of targeted lines of intervention during the lockdown would induce a further increase in poverty and inequality.

Another work deals with wealth distributions under the spread of infectious diseases\cite{dimarco2020wealth}. Considering the coupling of a compartmental epidemic dynamics with a kinetic model of wealth exchange, the authors found that that the spread of the disease seriously affects the distribution of wealth. Indeed, the evolution of disease together with the dynamics of wealth exchange changes the wealth distributions from a bimodal form to a fat-tailed one\cite{dimarco2020wealth}. Still talking about the economic implications of mobility restrictions, it was reported the decline of Gross Domestic Product in China\cite{huang2020quantifying}.

In this work, we discuss the effectiveness of social distance policies in developing and emerging countries where the share of informal employment is very high. Although it is not always true that there is a relationship between informal employment and poverty, we may find a clear positive relationship among them. It is worth mentioning that in developing and emerging countries the share of informal employment in total employment ranges from 50 $\%$ to more than 98 $\%$ \cite{ILO2018}. In this context, we investigate emerging scenarios for a generalized SIR-like model taking into account a heterogeneous propensity of individuals to comply with
the self-isolation policies. Our work differs from all the previously mentioned studies in the sense of that we split the population following economic attributes of the society. As we will see, the fraction of works in the informal sector plays a crucial role in the emergence or not of a second peak in the COVID-19 spreading.

Our work relates to the recent interesting contributions that consider the effect of social factors into epidemics models\cite{wang2015coupled,pires2017dynamics,2018piresOC,crokidakis2012probing} and works that have tried to study and forecast the early evolution of the COVID-19 pandemics and the public policy response to it\cite{CrokidakisRJ,bastos2020modeling,Kucharski2020,Berger2020,Read2020,walker2020}.  

The rest of the work is organized as follows: In Section \ref{sec:models}, we introduce the SIRASD models used in this work and how we split the population into two groups: the group of individuals that work in the informal economy can afford partially self-isolation and the rest of population that has the choice of self-isolation.
In Section \ref{sec:results}, we present the main results of our model. Furthermore, while Section \ref{sec:Discussion} presents a discussion of our results, Section \ref{Sec:Limitations} presents their limitations. Finally, Section \ref{sec:finalRemarks} stresses the main results of our work.

\section{Model}
\label{sec:models}

\begin{figure}[!hbtp]
\centering 

\includegraphics[width=0.45\linewidth]{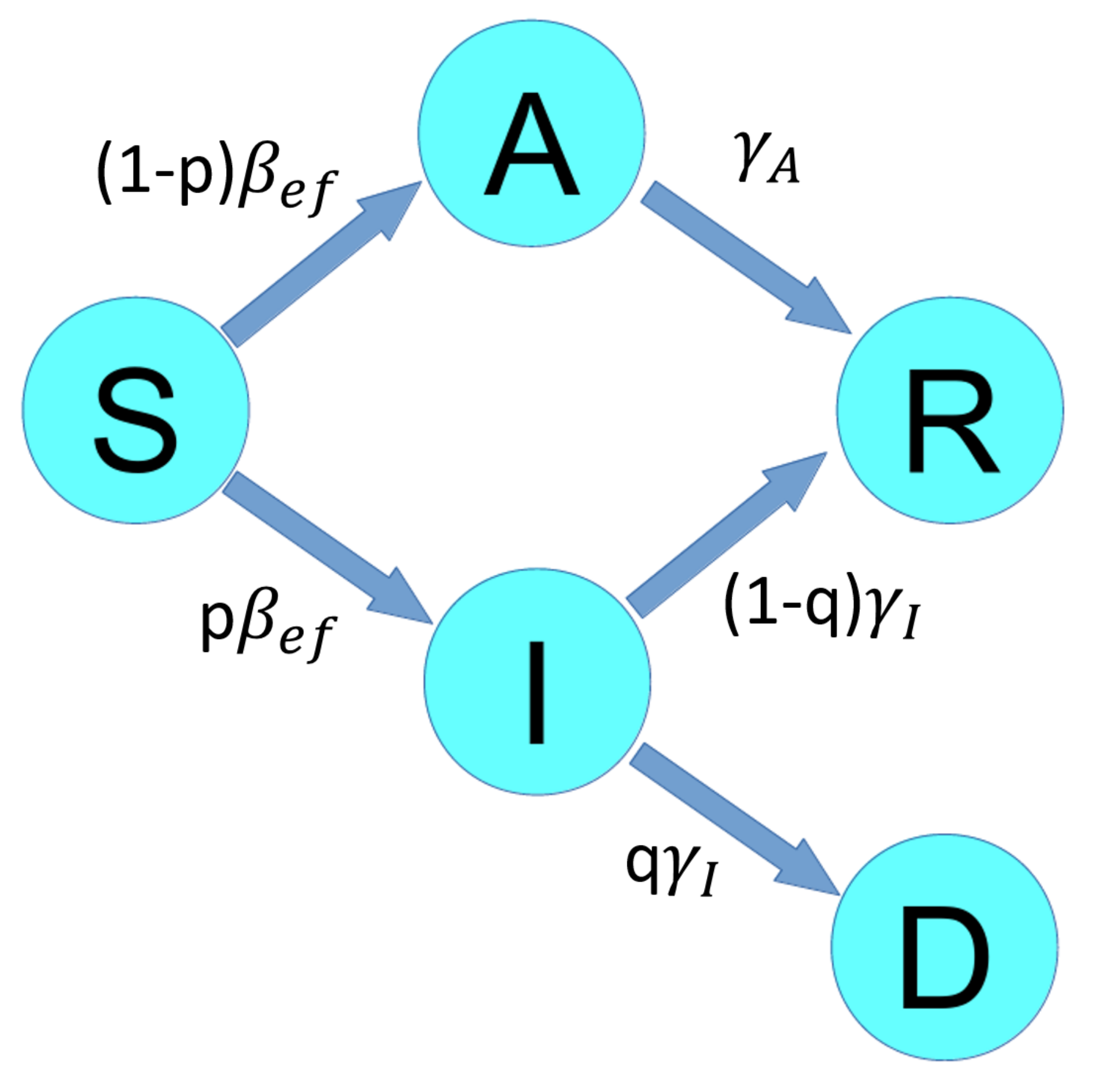}

\caption{Susceptible - Infected - Recovered - Asymptomatic - Symptomatic - Dead (SIRASD) compartmental model.}
\label{Fig:model}
\end{figure}

We divide the population into two types of individuals:

\begin{itemize}
\item Type 1: the group that has the option of self-isolation. This group represents a fraction $f_1$ of the full population. 
\item Type 2: low income workers  in the gig economy and informal sectors. This group  represents a fraction $f_2=1-f_1$ of the full population.
\end{itemize}

Let $\phi_u$ be the noncompliance degree of the group $u\,(u=\{1,2\}$) concerning governmental containment policies. Thus
$1-\phi_u$ is the degree of engagement with self-isolation advice.

For the COVID-19 there are both asymptomatic and symptomatic cases. 
Thereby we consider a framework close to a recent proposal\cite{bastos2020modeling}
 (and references therein). That is we consider a SIRASD 
 (Susceptible-Infected-Recovered-Asymptomatic-Symptomatic-Dead)
 model, where here extend it for the inclusion of two groups.

To  explain in detail our model consider two individuals $\{i,j\}$ belonging to the groups 
 $\{u,z\}$, respectively. Then
 
\begin{itemize}
\item If $i$ is in the state S
and if $j$ is infected in the state $X=\{A \text{ or } I\}$ then a transmission event occurs in which 
$i$ enters in the state I with rate $p\phi_z\phi_u\beta_X$ 
or 
enters in the state A with rate $(1-p)\phi_z\phi_u\beta_X$, where $p$ is the proportion of individuals who develop symptoms.

\item If $i$ is in the state A then it enters in the state R with rate $\gamma_A$.

\item If $i$ is in the state I then it enters 
in the state D with rate $q\gamma_I$, otherwise it enters in the state R with rate $(1-q)\gamma_I$. In such case, $q$ is the probability of an individual in the class I dying from infection before recovering.

It is important to stress that $D(t)$ informs how many individuals who tested positive for COVID-19 were declared dead at date $t$.
\end{itemize}

An illustration of the transition between the compartments is shown  in Fig.~\ref{Fig:model}. From the above-stated rules the set of coupled ODEs that govern the system considering the mean-field assumption. Explicitly, we arrive at:
\begin{equation}\label{Eq:Su}
 \frac{dS_u}{dt}  = -  \frac{S_u}{N}  \sum_{z=1}^{2} \phi_u\phi_z(\beta_II_z+\beta_AA_z), 
\end{equation}
\begin{equation}\label{Eq:Au}
 \frac{dA_u}{dt}  = \frac{S_u}{N} (1-p) \sum_{z=1}^{2} \phi_u\phi_z(\beta_II_z+\beta_AA_z)  - \gamma_A A_u,
\end{equation}
\begin{equation}\label{Eq:Iu}
 \frac{dI_u}{dt}  = \frac{S_u}{N} p\sum_{z=1}^{2} \phi_u\phi_z(\beta_II_z+\beta_AA_z)  - \gamma_I I_u,
\end{equation}
\begin{equation}\label{Eq:Ru}
 \frac{dR_u}{dt}  = (1-q)\gamma_I I_u +  \gamma_A A_u,
\end{equation}
\begin{equation}\label{Eq:Du}
 \frac{dD_u}{dt}  = q\gamma_I I_u, 
\end{equation}
where $N=\sum_{u=1}^{2}(S_u+A_u+I_u+R_u)$. From the aforementioned equations we define the effective transmission rate that is shown in Fig.\ref{Fig:model},
\begin{equation}\label{Eq:beta_ef}
 \beta_{ef}^u =  \sum_{z=1}^{2} \phi_u\phi_z(\beta_I\frac{I_z}{N}+\beta_A\frac{A_z}{N}) ~,
\end{equation}
where the terms $\phi_u\phi_z$ show that the interaction  can involve individuals within the same group (intragroup interaction: $\phi_1\phi_1$, $\phi_2\phi_2$) or between different groups (intergroup interaction: $\phi_1\phi_2$, $\phi_2\phi_1$).


We intend to model scenarios that arise, as above-mentioned, in emerging and developing economies, where the number of individuals in the informal economy is a large stake of the total labor force. In order to provide convincing numerical arguments, our model uses epidemiological parameters that come directly or indirectly from  Ref.~\cite{bastos2020modeling}, that were estimated from the COVID-19 pandemics that has taken place in Brazil:
$\beta_A=0.458$, $\beta_I=0.455$, 
$\gamma_A=0.144$,
$p=0.624$, $q=0.029$, $\gamma_I=0.149$ and $\phi_u=0.799$.
For further comments on $q$, $\gamma_I$ and $\phi_u$ see the Appendix A. Here we consider $N=210147125$ as the total population (similar to Brazil).
We consider an initial condition as $I_1(t_0)=1$ and  $A_1(t_0)=0.5$ for the group $1$. For the group $2$ we set $I_2(t_0)=A_2(t_0)=0$.

\section{Results}
\label{sec:results}

\begin{figure*}[!hbtp]
\centering 
\includegraphics[width=0.99\linewidth]{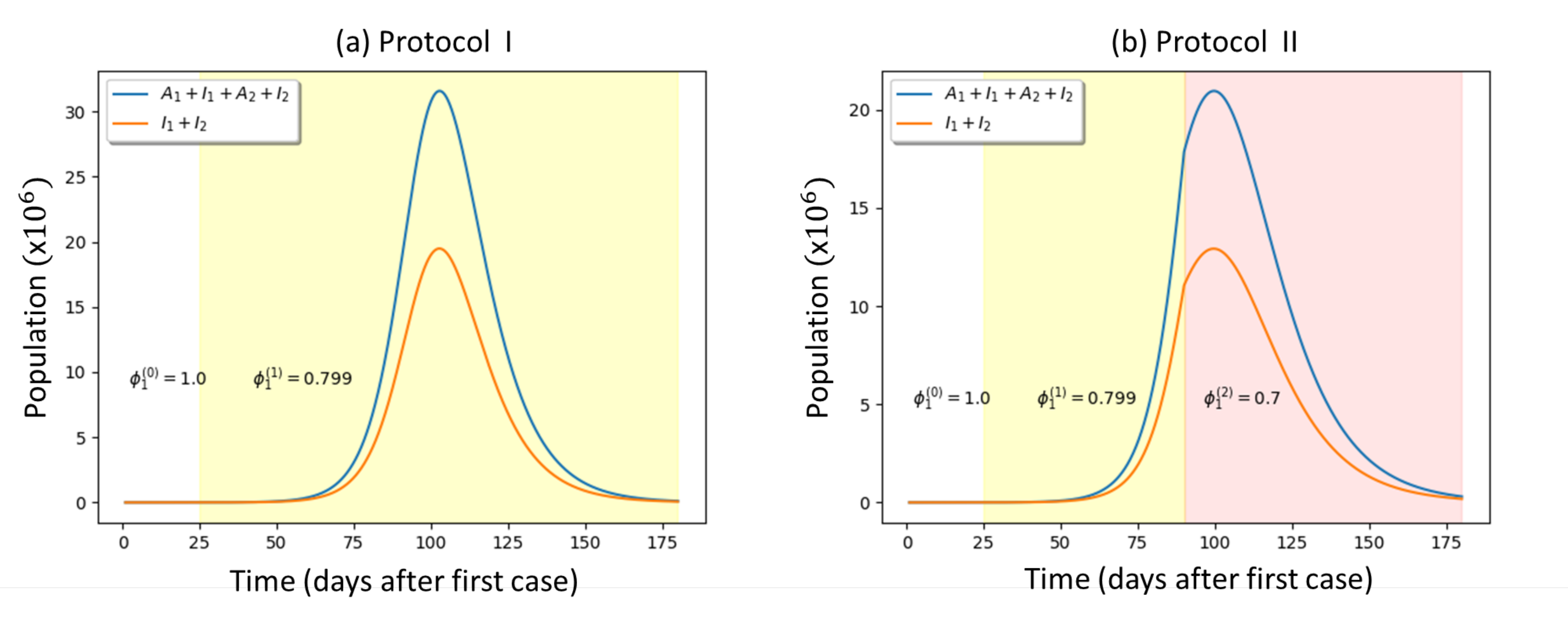}

\caption{Time series for the number of individuals in the class  $\sum_i I_i$ as well as $\sum_i (A_i+I_i)$ considering: (a)  protocol I (b) protocol  II. In the protocol II we apply $\phi=0.799 \rightarrow \phi=0.7$ on day $t_{policy}^{(2)}=90$ after the first case (red shaded region).}
\label{Fig:time-series-br-1}

\vspace{1cm}

\centering 

\includegraphics[width=0.99\linewidth]{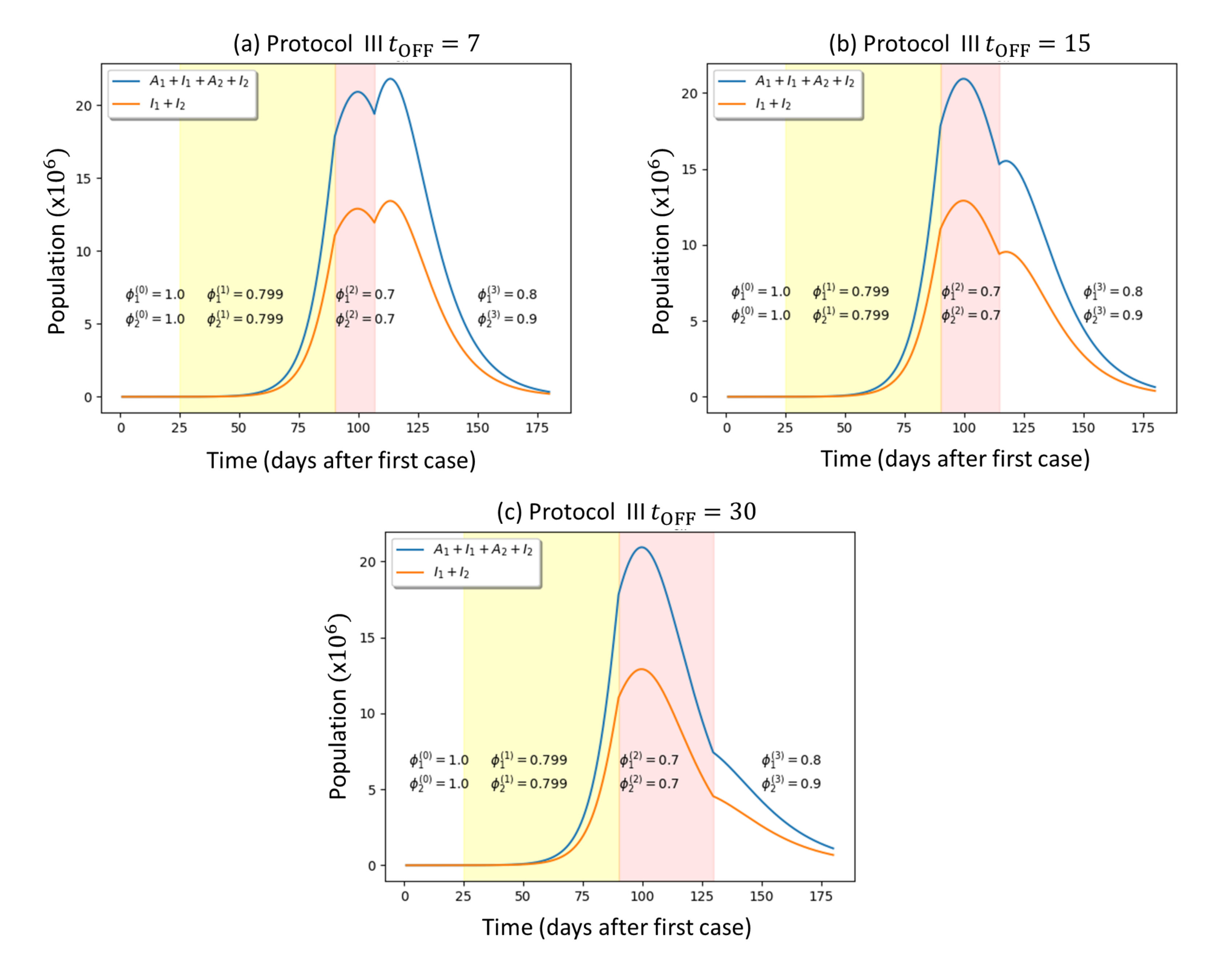}

\caption{Time series for the number of individuals in the class $\sum_i I_i$ as well as $\sum_i (A_i+I_i)$ considering: (a)  $t_{\rm OFF}=7$, (b) $t_{\rm OFF}=15$ and (c)  $t_{\rm OFF}=30$. The first white, yellow and red shaded areas are explained in the previous Figure. The last white region represents the case with soft self-isolation rules.}
\label{Fig:time-series-br-2}
\end{figure*}


\begin{figure*}[!hbtp]
\centering 

\includegraphics[width=0.61\linewidth]{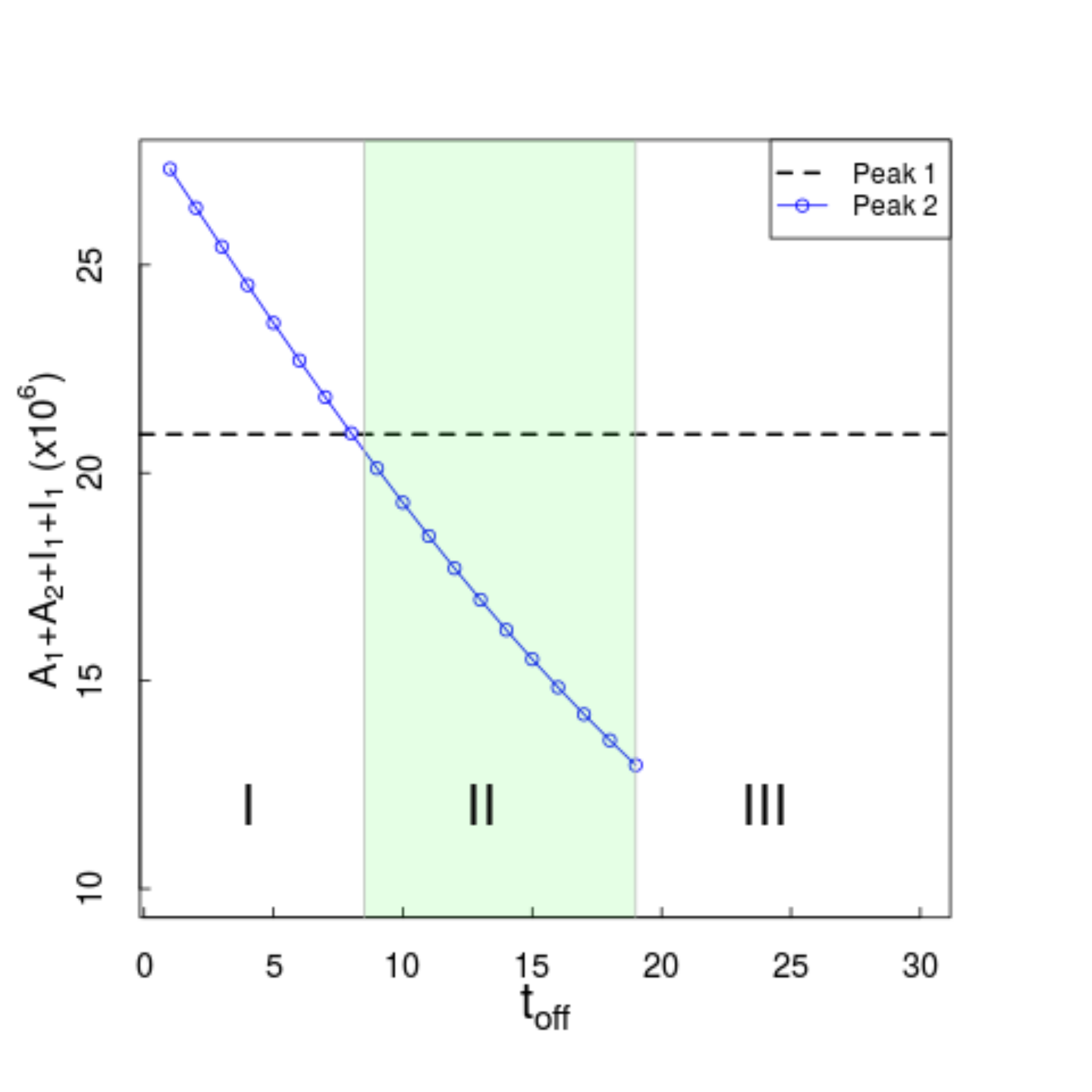}

\caption{Dependence of  the peak size of $\sum_i (A_i+I_i)$ with $t_{\rm OFF}$. Parameters: $t_{max}=365$,  $f_1=0.6$, $\phi_1^{(2)}=\phi_2^{(2)}=0.7$, $\phi_1^{(3)}=0.8$ and $\phi_2^{(3)}=0.9$. Regime I: the second peak is larger than the first one. Regime II: the secondary peak is smaller than the first one. Regime III:  absence of a second peak. Each of these regimes is illustrated in Fig. \ref{Fig:time-series-br-2}.} 
\label{Fig:peaks-res-toff-br}

\vspace{1cm}

\centering 

\includegraphics[width=0.99\linewidth]{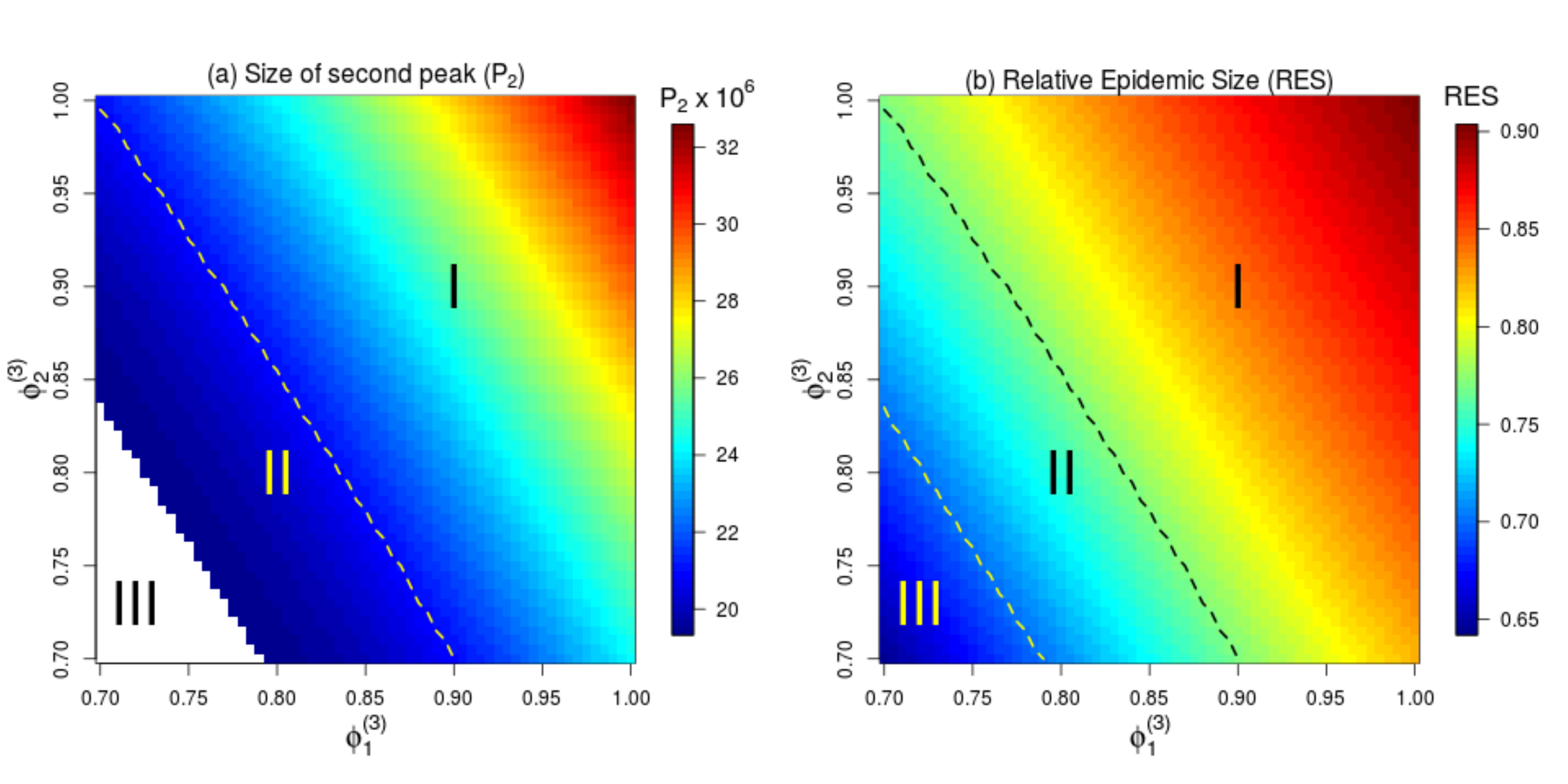}

\caption{Diagrams $\phi_1^{(3)}$ vs $\phi_2^{(3)}$ for: (a) $P_2$ and (b) RES. Results  obtained for $t_{max}=365$ days, $t_{\rm OFF}=7$,  $f_1=0.6$ and $f_2=0.4$. The regimes I,II and III are explained in the Fig.\ref{Fig:peaks-res-toff-br}.  $P_2$ is computed considering both symptomatic and asymptomatic individuals, ie $A_1+A_2+I_1+I_2$.}
\label{Fig:peaks-phioff-br}
\end{figure*}


In this section we present the results solving our coupled ODEs using the \emph{solveivp} of python.
Specifically, we use the  \emph{RK45} method that implements an explicit Runge-Kutta method of order 5(4). Such procedure manages the error considering an accuracy of the 4-order and it employs a 5-order accurate formula to take the steps.  

It is important to mention that second waves of infections can be observed measuring distinct quantities, like the new daily cases or the number of active cases in a given day\cite{leung2020first,castro2020second,sansao2020second,faranda2020modelling,vaid2020risk,xu2020beware,netoactive,singh2020predictive,menon2020modelling,GhanbariIran}. In this work, we chose to exhibit the number of active cases in a given day.

Apart from the number of individuals in each class, there is a second quantity of interest, namely the 
Relative Epidemic Size (RES) that is computed from $t_0$ to $t$
\begin{equation}\label{Eq:res}
RES = \sum_{z=1}^{2}
\frac{S_z(t_0) - S_z(t)}{N} ~.
\end{equation}

\noindent
In order to better grasp our full protocol lets first consider the case with $f_1=1$.
 Let $u$ be the index of group $u$. We consider $\phi_u=\phi_u^{(0)}=1$ during the initial stage of the epidemic spreading because the level of self-isolation is almost null.
 We shall assume $\phi_u^{(0)} \rightarrow \phi_u^{(1)}=0.799$ on day $t_{policy}^{(1)}=25$  after the beginning of the epidemic spreading. With this procedure (we call it  protocol I) we obtain the time series shown in Fig.\ref{Fig:time-series-br-1}(a) that recover the results presented in Ref.~\cite{bastos2020modeling} considering the scenarios with the current confinement rules imposed by the government for an indefinite time. Taking into consideration that the value of the total population is 210 million people, the peaks in the panels are between $9\%$ and $14\%$ for Infected+Asymptomatic cases and circa $5\%$ for the Infected cases alone. In spite of the subnotification issues that have been reported\cite{volpattoavaliaccao,paixo2020estimation,covid19brazil}, these figures are compatible with the fraction of infected people computed in other countries close to $10\%$ as well
~\cite{flaxman2020estimating}.

Consider the protocol II shown in Fig.\ref{Fig:time-series-br-1}(b). During the explosive growth of the epidemic, the isolation policy is improved by better surveillance. Explicitly, we decrease the  noncompliance degree from $\phi_u^{(1)}=0.799$ to  $\phi_u=\phi_u^{(2)}$ on day $t_{policy}^{(2)}=90$ after the first case at day $t_0$.
Henceforth we set $\phi_u^{(2)}=0.7$, but the nature of our results does not change qualitatively for other values, as discussed in the Appendix C. In Fig.~\ref{Fig:time-series-br-1}(b), we see the strengthening of the confinement restrictions leads to a substantial decrease in the number of symptomatic and asymptomatic individuals.

The self-isolation measures are permanent in the protocols I and II. However, after the  epidemic growing phase, 
there might be political and economic pressure to ease strict confinement rules. In that sense, let us move to the protocol III with temporary self-isolation guidelines. Explicitly,

\begin{itemize}
\item After each time step (day) we monitor $\delta I(t) = \sum_z \left( I_z(t) - I_z(t-1)  \right) $.
\item At $t_0$ we set $t_{decrease}=0$. For each $dI(t)<0$ we increase $t_{decrease}$ in one unit. 
\item If $t_{decrease}=t_{\rm OFF}$  we  set $\phi_u=\phi_u^{(3)}$. That is if $dI(t)<0$ during $t_{\rm OFF}$ consecutive days, the social distancing rules are relaxed. 
\end{itemize}

As one can see in the above items, the method we consider is applicable after the number of active cases has reached a peak. Figure~\ref{Fig:time-series-br-2} exhibits the time series for the number of individuals infected considering $f_1=0.6$ and $f_2=0.4$. We have made this choice due to a recent poll in Brazil made by the Brazilian Institute of Geography and Statistics (IBGE), that stated $39.9\%$ of the population works in the informal economy\cite{ibge}, which leads to $f_2=0.4$. However, we considered other values of $f_1$ and $f_2$ in Appendix C. The self-isolation measures are lifted  $t_{\rm OFF}$ days after the peak. At that moment, the degree of noncompliance is increased to $\phi_1^{(3)}=0.8$ 
and $\phi_2^{(3)}=0.9$ (last white regions in Fig. \ref{Fig:time-series-br-2}). 
If the interruption of the confinement rules takes place one week after the peak, 
$t_{\rm OFF}=7$, we see that the second outbreak is larger than the first one. This scenario is different for $t_{\rm OFF}=15$, where the secondary peak is smaller than the first one. If $t_{\rm OFF}=30$ days, then there is no rising of the secondary peak even though there is a rise in the person-to-person contagion.

Figure~\ref{Fig:peaks-res-toff-br} shows how the time for interruption of the confinement rules impacts the epidemic spreading behavior. The peak size is computed taking into account both symptomatic and asymptomatic individuals $A_1+A_2+I_1+I_2$. Specifically, there are three main outcomes. Easing the mobility restrictions too soon triggers an abrupt rise of the new cases
that leads to a pronounced second peak that is worse than the first one. This is the regime I. In regime II, the secondary chain of contagion also leads to a new noticeable outbreak but now with magnitude  smaller than the first one. In regime III, there is no second local maximum.
Then, we highlight that there are two thresholds: (i) for prevention of a second large-scale epidemic outbreak;  (ii) for  prevention of a second small-scale outbreak. 

Figure~\ref{Fig:peaks-phioff-br} disentangles the role played by the degree of noncompliance $\phi_u^{(3)}$ of each group $u$. When the confinement guidelines are lifted too early ($t_{\rm OFF}=7$) the majority of the combinations of $\phi_1^{(3)}$ vs  $\phi_2^{(3)}$ leads to the regime I where the second outbreak is more aggressive than the first one. In this setting, the relative epidemic size (RES) can achieve about ~$90\%$ of the population in the long-run (1 year in such figure). For combinations of moderated values of $\phi_1^{(3)}$ vs  $\phi_2^{(3)}$, there is a substantial region in regime II where RES is mostly between $70\%$-$80\%$ of the population. The non-negligible presence of the regime III indicates that  the prevention of a secondary epidemic outbreak can be achieved if the  engagement of the population with the stay-at-home guidelines does not decrease too much.

\begin{figure*}[!hbtp]
\centering 
\includegraphics[width=0.995\linewidth]{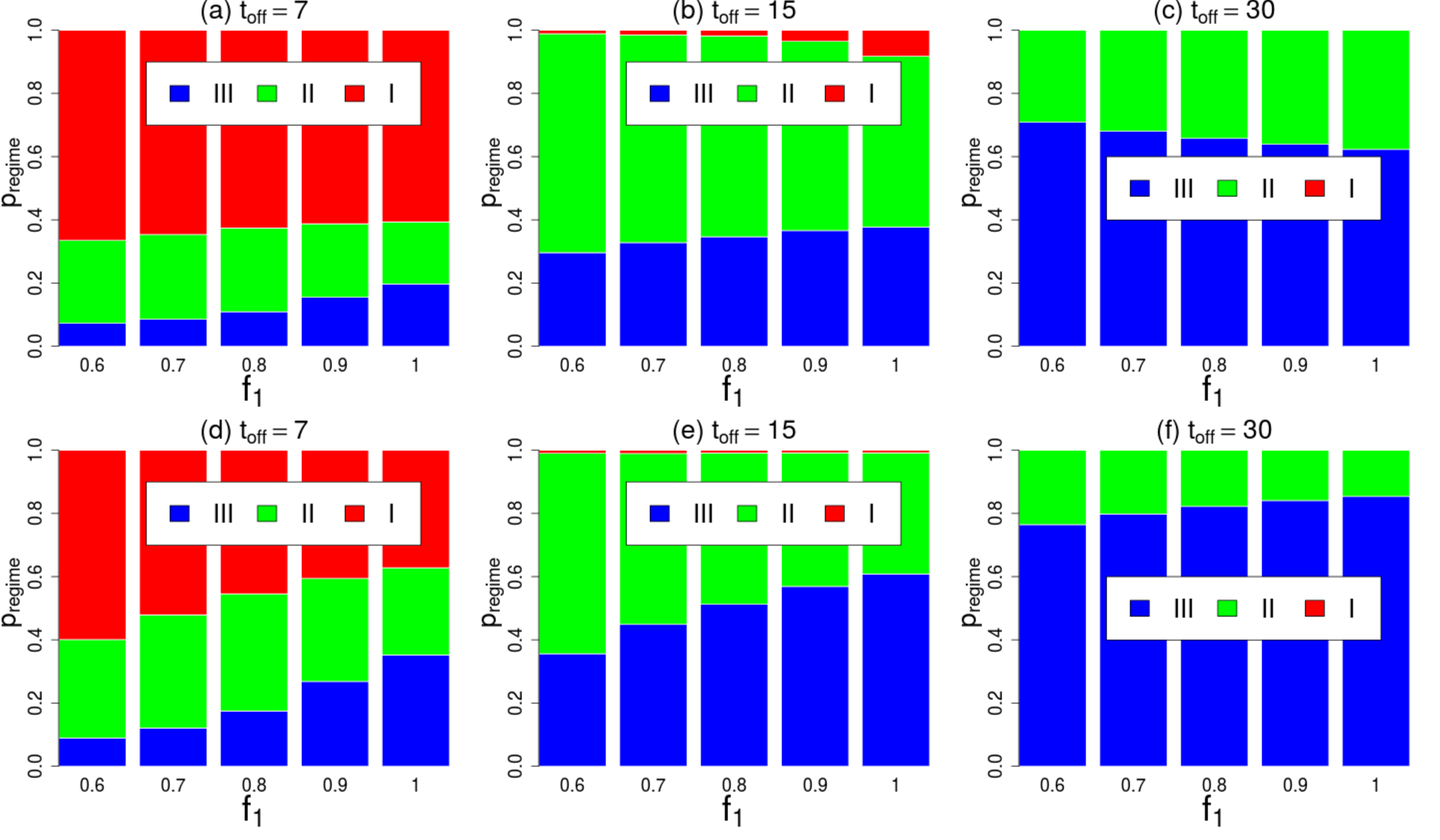}

\caption{Barplot with the proportion  of each regime $p_{regime}$ in diagrams similar to the shown in Fig.\ref{Fig:peaks-phioff-br}. (Top) All $61x61$ combinations of  $\phi_1^{(3)} \times \phi_2^{(3)} \in [0.7,1] \times [0.7,1]$. (Bottom) Combinations satisfying $\phi_2^{(3)}\geq \phi_1^{(3)}$. Regime I: the second peak is larger than the first one. Regime II: the secondary peak is smaller than the first one. Regime III:  absence of a second peak. Outcomes for: (a,d)  $t_{\rm OFF}=7$, (b,e) $t_{\rm OFF}=15$ and (c,f)  $t_{\rm OFF}=30$.}
\label{Fig:barplot-br-1}
\end{figure*}


\begin{figure*}[!hbtp]
\includegraphics[width=0.995\linewidth]{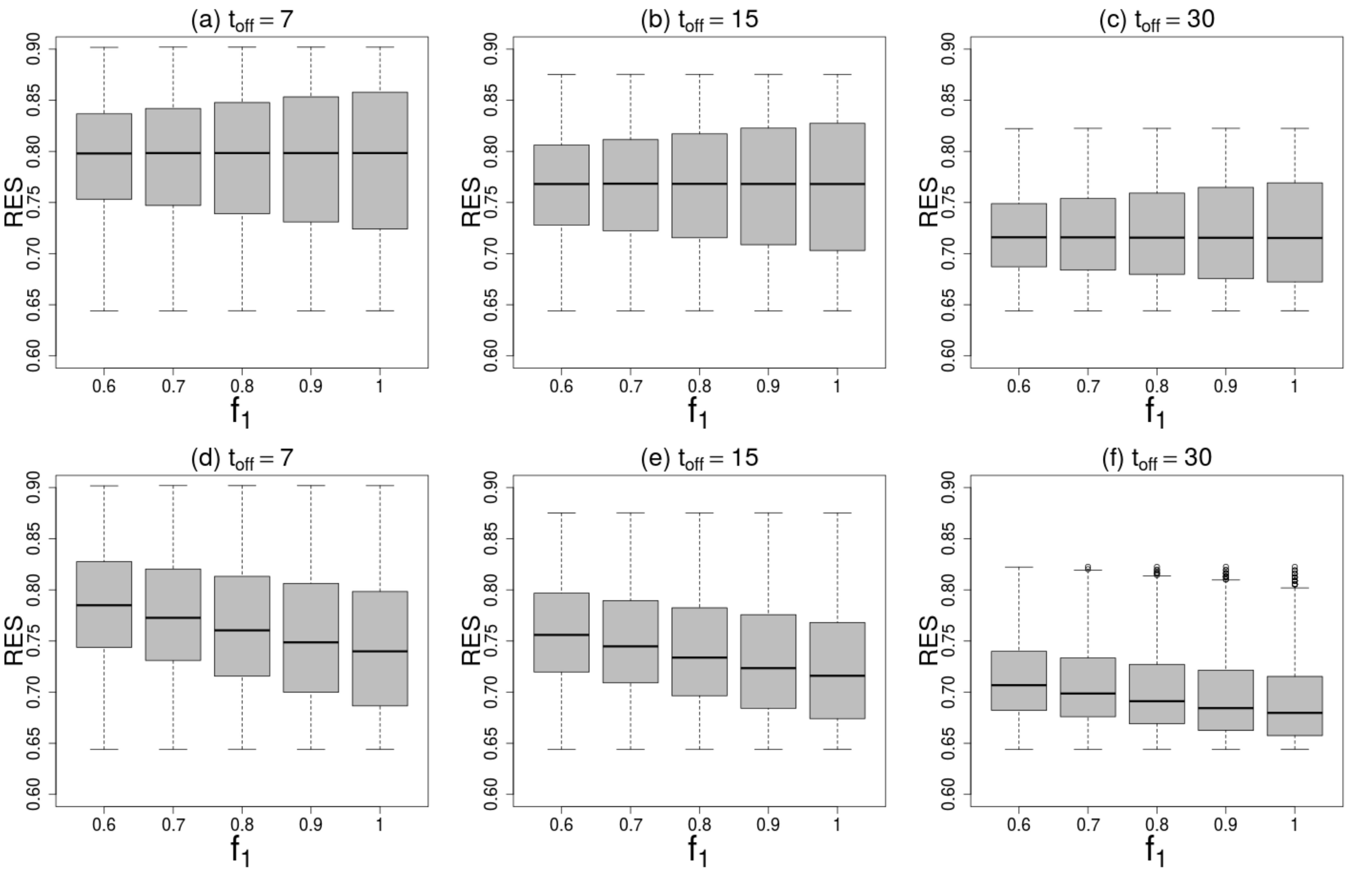}

\caption{Boxplot with the range of values exhibited by RES in diagrams similar to the shown in Fig.\ref{Fig:peaks-phioff-br}.  (Top) All $61x61$ combinations of  $\phi_1^{(3)} \times \phi_2^{(3)} \in [0.7,1] \times [0.7,1]$. (Bottom) Combinations satisfying $\phi_2^{(3)}\geq \phi_1^{(3)}$. Results for: (a,d)  $t_{\rm OFF}=7$, (b,e) $t_{\rm OFF}=15$ and (c,f)  $t_{\rm OFF}=30$.}
\label{Fig:boxplot-br-1}

\end{figure*}

Let us now turn our attention to the main results depicted in  Figs.\ref{Fig:barplot-br-1}-\ref{Fig:boxplot-br-1}
for $f_1=\{0.6,...,1\}$ and $t_{\rm OFF}=\{7,15,30\}$. 
 In panels (a-c) each barplot or boxplot is obtained considering grids with $61\times 61$ combinations of  $\phi_1^{(3)} \times \phi_2^{(3)} \in [\phi_1^{(2)},1] \times [\phi_2^{(2)},1]$ where $\phi_1^{(2)}=\phi_2^{(2)}=0.7$. Thus, all the panels (a-c) totalize $ 3\times 5\times 61\times 61 = 55815$  different projections. The panels (d-f) show the results for the those combinations satisfying $\phi_2^{(3)}\geq \phi_1^{(3)}$. In the boxplot the gray shaded box goes from the first quartile to the third quartile and the horizontal line inside the box is the median.

Figure~\ref{Fig:barplot-br-1} shows the barplots for the proportion  of each regime $p_{regime}$ for several $f_1$ and $t_{\rm OFF}$. In the setting with $t_{\rm OFF}=7$ and $f_1=0.6$, the overwhelming majority of configurations lead to the establishment of the regime I, as previously observed. But, this advantage of the regime I decrease as $f_2$ decreases (by increasing $f_1$). 
In the setting with $t_{\rm OFF}=15$ all the scenarios exhibit a smaller proportion for the regime I in comparison with corresponding scenarios for $t_{\rm OFF}=7$.
However, there is a dual effect of rising $f_1$. On the one hand, it increases the proportion of configurations associated with the regime III. On the other hand, it also increases the possibilities for the emergence of regime I. In the setting with $t_{\rm OFF}=30$ we also see a double-edged sword: (a) the percentage of regime I is null and all the percentage of the regime III is higher than the corresponding to the cases $t_{\rm OFF}=\{7,15\}$; (b) an increase of $f_1$ increases the relative advantage of regime II. These nonmonotonic effects arise because some combinations  $\phi_1^{(3)} \times \phi_2^{(3)}$ favor the regime $I$ and other combinations favor the regime $III$ 
as depicted in Figure~\ref{Fig:peaks-phioff-br}.
Such mechanism is corroborated with the panels (d-f)
where we see that the combinations satisfying
$\phi_2^{(3)}\geq \phi_1^{(3)}$ leads to a monotonic behavior
 of $p_{regime}$ vs $f_1$ for all  $t_{\rm OFF}=\{7,15,30\}$.

Figure~\ref{Fig:boxplot-br-1} shows the boxplots for RES considering decreasing values of $f_2=1-f_1$ as well for increasing values of $t_{\rm OFF}$. Such results show that an increment in 
$t_{\rm OFF}$ leads to an overall decrease in the relative epidemic size (RES). But a detailed analysis in
each panel shows that an increase in $f_1$ produces an increase in the  interquartile range  of values for RES (gray area). This indicates the presence of a twofold effect since RES can achieve smaller values as $f_1$ increases, but it also leads to the possibility for RES reaching higher values. Again such twofold effect arises because some combinations  $\phi_1^{(3)} \times \phi_2^{(3)}$ are responsive for an increase in RES and other combinations promote a decrease in RES as unveiled in Figure~\ref{Fig:peaks-phioff-br}. This is confirmed with the panels (d-f) where the combinations satisfying $\phi_2^{(3)}\geq \phi_1^{(3)}$  
leads to a decrease in RES as 
 $f_1$ increases
 for all  $t_{\rm OFF}=\{7,15,30\}$.

\section{Discussion}
\label{sec:Discussion}

The findings in Figs. \ref{Fig:barplot-br-1}-\ref{Fig:boxplot-br-1} are our main results. 
Such figures show that, for our parameters, 
it is very likely the emergence of a second peak (regimes I+II) if the preventive measures are lifted too soon. Even more alarming, there is a non-negligible risk for the magnitude of such second peak be higher than the first one (regime I).  Apart from this, we note that for a given $t_{\rm OFF}$  there is the  possibility for a twofold effect in which an intervention designed to hamper the epidemic spreading can backfire.
However, in such a situation the  establishment of positive or negative outcomes depends on the combinations of $\phi_1^{(3)}$ vs $\phi_2^{(3)}$ as indicated in Fig.\ref{Fig:peaks-phioff-br}.
Such findings highlight that it is significant to have a substantial alignment between different interventions designed to decrease the degree of noncompliance as well as to support the fraction of the population that cannot afford for the self-isolation  even after the first peak of spreading. Moreover, complementary studies using different parameters we could verify that the present model is also capable of reproducing different situations of separated peaks as found in several U.S.A. cities during the Spanish  flu pandemics\cite{bootsma2007effect,hatchett2007public}. Therein, it is possible to assess the impact of different public health measures in the number and evolution of fatalities, with some cities basically exhibiting a single peak (an indicator of proper policies)  and other cities with significant second peaks. Importantly, some of the cities showing two peaks were cities that have not had good governance and provided adequate responses to the COVID-19 pandemics. In other words, although we have adjusted our model to the present COVID-19 case, our model is likely to be relevant, in theoretical viewpoint, in the analysis of other situations, namely the computational forward testing of public health policies.

Other correlated works considering 
COVID-19 spreading  have also shown the  possibility of a second epidemic peak. In Ref.~\cite{rogers2020ending}, it is shown -- 
with variants of the SIR model -- the potential of the second peak
of infections for the UK. In Ref.~\cite{hoertel2020lockdown},
the authors calibrated a stochastic agent-based model from data in France and they projected that it would be unlikely to prevent 
the second chain of contagions once quarantine is lifted. A second chain of spreading was also predicted -- using a generalization of
the SIR model -- as a potential outcome for Italy after the 
relaxation of the mobility  
restrictions\cite{pedersenquantifying}. 
A recent work considering the case of Brazil in a group-free  
Susceptible-Exposed-Infected-Recovered-Dead model presented some 
time series suggesting that  the social isolation must hold until 
the end of 2020 in order to diminish the chance of the second peak
\cite{cintra2020estimative}. Effectively, the conclusion of all those works is that the safer situation is to hold the isolation 
for as long as possible in order to decrease the magnitude of the second peak. 

Our results reinforce the importance of the need to keep the social isolation as long as possible or at least to avoid the activities that cause the most crowds of the population, including indoor activities and private parties (since the public ones are forbidden). Furthermore, the ministry of health should stress  the unconditional use of protective masks. In addition, the government must negotiate the vaccines that are available on the market and plan their distribution in the vast Brazilian territory, since this is the only measure that is able to protect the population without affecting the economy.

For further discussion  see our Appendices.

\section{Limitations}
\label{Sec:Limitations}

We consider that as the epidemic starts to climb sharply there will be an increased  pressure to decrease the degree of noncompliance (red shaded region in Fig.\ref{Fig:time-series-br-2}). At this point we still assumed the same level of compliance of both groups because of the current implementation of income transfer for the group $2$\cite{auxilio}. After the first peak and as soon as the stay-at-home restrictions are suspended we set different levels of compliance with the post-quarantine stage for each group (last white  region in Fig.\ref{Fig:time-series-br-2}).

Besides, our work does not consider explicitly an upper bound for the capacity of the healthcare system. Underreporting is another feature that is not modeled here and we have not considered the clear regional heterogeneity in Brazil as well.

We have also assumed permanent immunity, i.e., after recovering from COVID-19 individuals cannot be infected anymore. In addition, we considered a mean-field-like approach, where each individual can interact will all others. In this case, spatial features were not considered in the model.

Although our work presents limitations in several dimensions, such as the level of compliance of the individuals, an unbounded healthcare system, the absence of underreporting, spatial homogeneity and permanent immunity, as above mentioned, the qualitative results of our work should remain. The level of compliance of the individuals seems to be the strongest assumption. For this reason, we have provided several variations of the parameters associated with this assumption. While the other assumptions may quantitatively affect  the results, they do not seem to change our qualitative results. If the number of infected people, that needs treatment, is larger than the number of available beds, the number of deaths will be larger than our model predicts\cite{NORONHA2020}. Underreporting may cause a nonlinear effect in the model\cite{bastos2020covid19}. There are two types of underreporting. The first situation happens when the underreported cases are related to the number of infected individuals. In general, larger underreporting means a reduction in the number of strong symptomatic infections, that implies that the disease is less dangerous than predicted by our model. The second situation happens when the number of deaths is underreported. In this case, the disease is more dangerous than predicted by our model. The lack of spatial homogeneity may cause variations in the speed of the spread of the pandemics. While the spread of the disease is faster in denser areas than in sparser areas, there is no evidence that they are deadlier in the denser ones. In fact, it may depend on the public policy choices of those areas\cite{carozzi2020}. Anyway, the qualitative results of our paper are also valid.

\FloatBarrier
\section{Conclusion}
\label{sec:finalRemarks}

Our work has investigated the evolution of COVID-19 pandemics in emerging and developing economies where the informal economy represents a large fraction of the total economy. 
The stratification in groups according to economic-based features is a novelty of our work. Although we have used parameters estimated for the Brazilian case to analyze the effectiveness of social distancing policies and to estimate the likelihood of arising a second peak, our results are qualitatively the same for all economies that show these characteristics. We apply a SIRASD model considering a  population split into two groups with different behaviors, namely a group that belongs to a class that is able to self-isolate and a group that is formed by low-income workers in the gig economy or informal sectors. While the first group usually belongs to the higher income class or is able to work at home, the second group is usually in a low-income class and supplies services to consumers and businesses, and is not able to provide their services in home office. In this context, the results show that the existence of these two types of social behaviors strongly affects the dynamics and possibility of a second peak in the evolution of COVID-19.  Based on these results, it is possible to understand that in order to   curb the pandemic, it is fundamental the adherence of low-income people --- who largely make their living on informality --- in the self-isolation policies as pointed by public health authorities worldwide. In order to solve the dilemma choosing between i) going out to get few earnings and risk being infected or ii) stay home and face starvation in favor of the latter, the present results signal it is pivotal the design of income transfer policies that pay for these people to stay at home at least 30 days after of the first peak.

\section*{Acknowledgments}
The authors acknowledge financial support from the Brazilian funding agencies CNPq, CAPES and FAPERJ.

\newpage

\appendix

\section{\label{sec:disApp}Epidemiological parameters}

Using the acronyms presented in the main text, we first present the model
recently proposed\cite{bastos2020modeling}:
%
\begin{equation}\label{Eq:Su_caj}
 \frac{dS}{dt}  = -  \frac{S}{N}  \psi(\beta_II+\beta_AA), 
\end{equation}
\begin{equation}\label{Eq:Au_caj}
 \frac{dA}{dt}  = \frac{S}{N} (1-p) \psi(\beta_II+\beta_AA)  - \gamma_A A,
\end{equation}
\begin{equation}\label{Eq:Iu_caj}
 \frac{dI}{dt}  = \frac{S}{N} p \psi(\beta_II+\beta_AA)  - \frac{\gamma_S}{1-\rho} I,
\end{equation}
\begin{equation}\label{Eq:Ru_caj}
 \frac{dR}{dt}  = \gamma_S  I +  \gamma_A A,
\end{equation}
\begin{equation}\label{Eq:Du_caj}
 \frac{dD}{dt}  =  \rho\frac{\gamma_S}{1-\rho}  I, 
\end{equation}

Our stratified-population model considering two groups $u=\{1,2\}$ is given by Eqs.~(\ref{Eq:Su})-(\ref{Eq:Du}). To obtain $q$, $\gamma_I$ and $\phi_u$ we still need to apply a term-by-term comparison between the group-free model in Eqs.~(\ref{Eq:Su_caj})-(\ref{Eq:Du_caj}) and our group-based model in Eqs.~(\ref{Eq:Su})-(\ref{Eq:Du}):
\begin{itemize}
\item $q\gamma_I \equiv \gamma_S \rho/(1-\rho)$
\item $(1-q)\gamma_I \equiv \gamma_S$
\item $\phi_u\phi_u=\psi$. 
\end{itemize}
Thus,  $q=\rho=0.029$, $\gamma_I=\gamma_S/(1-\rho)=0.145/0.971=0.149$ and $\phi_u=\sqrt{\psi}=\sqrt{0.638}=0.799$ as described in the main text.

\section{\label{sec:repnum}Reproductive number analysis}

\begin{figure*}[!hbtp]
\centering  
\includegraphics[width=0.9\linewidth]{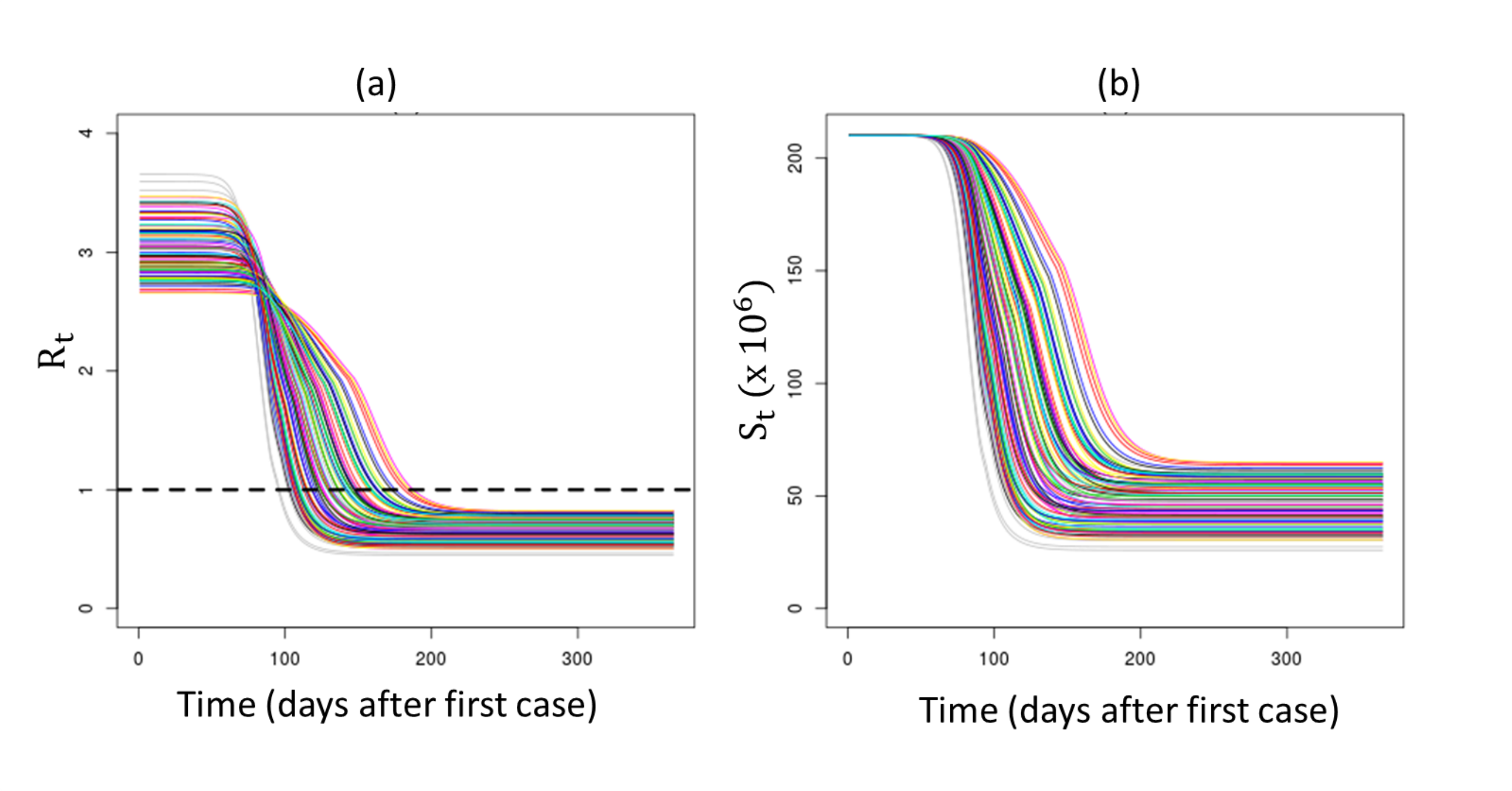}
\caption{Representative time series for $R=s(t)R_0$ and $S_t$. Other parameters: $f_1=0.6$, $t_{off}=7$, $\phi_u^{(0)}=1$, $\phi_u^{(1)}=0.799$, $\phi_u^{(2)}=0.7$,  $\phi_1^{(3)}=0.8$ and $\phi_2^{(3)}=0.9$.  } 
\label{Fig:Rt}
\end{figure*}

A common approach to estimate the basic reproductive number $R_0$ consists in estimating first the initial exponential growth rate characteristic of most human infectious diseases of rapid dissemination. Then $R_0$ can be estimated by a linearization of the models' equations (2) and (3) for small times\cite{anderson1992infectious,burr2006,brauer2006some}. In such case, considering the initial conditions $I_1(0)=1$, $A_1(0)=0.5$, $R_1(0)=0$, $D_1(0)=0$,
$S_1(0)=N\,f_1-I_1(0)-A_1(0)\simeq N\,f_1$ as well as
 $I_2(0)=0$, $A_2(0)=0$, $R_2(0)=0$, $D_2(0)=0$,
 $S_2(0)=N\,f_2$
 we can obtain $R_0 = R_0^{A_1}+R_0^{A_2}+R_0^{I_1}+R_0^{I_2}$, where
\begin{eqnarray}
R_0^{A_u} & = & \frac{f_u(1-p)(\phi_{u}^{(0)})^{2}\beta_A}{\gamma_A} \\
R_0^{I_u} & = & \frac{f_{u}p(\phi_{u}^{(0)})^{2}\beta_I}{\gamma_I}
\end{eqnarray}
for $u=1,2$. Taking into account the parameters considered in the text, i.e., $f_1=0.6$, $f_2=0.4$, $p=0.624$, $\phi_{1}^{(0)}=\phi_{2}^{(0)}=1$, $\beta_{A}=0.458$, $\beta_{I}=0.455$, $\gamma_{A}=0.144$ and $\gamma_{I}=0.149$, we obtain for the initial evolution of COVID-19 in Brazil
\begin{equation}
R_{0} = 3.1    
\end{equation}

As discussed in the literature\cite{burr2006}, while the basic reproductive number is more relevant for the case of emerging infectious diseases, in the case of endemic or recurrent infectious diseases, the reproductive number $R_t$ is a more practical quantity because it accounts for the residual immunity in the population due to previous exposures or isolation/vaccination campaigns in the population. Hence, we can write $R_t=R_0\,s(t)$, where $s(t)$ is the proportion of the total population that is effectively susceptible at time step $t$. Once the effects of isolation policies begin to take hold and the number of susceptible individuals decreases, the reproductive number $R_t$ will decay. Hence, the goal of public health authorities is to bring the reproductive number to $R_t<1$ as soon as possible.

In Fig. \ref{Fig:Rt} we exhibit the time evolution of the reproductive number $R_t$  and the total number of Susceptibles $S_t$ as functions of time, considering the parameters we used in the main text, i.e., the above-mentioned parameters and also $\phi_1^{(1)}=\phi_2^{(1)}=0.799$, $\phi_1^{(2)}=\phi_2^{(2)}=0.7$,  $\phi_1^{(3)}=0.8$ and $\phi_2^{(3)}=0.9$. We consider the 
outcomes for 100 curves using parameters randomly sampled in the confidence interval provided previously\cite{bastos2020modeling}.  
One can see that $R_t$ decreases after the implementation of isolation policies. As discussed in the main text, the noncompliance degree $\phi$ decreases at at $t=90$, which reflects in $R_t$ as a rapid decrease of the reproductive number. As Brazil did not adopted a lockdown, there is no jump on $R_t$ as observed in some European countries\cite{flaxman2020estimating}. However, we expected the desired behavior after some time, i.e., we will observed $R_t<1$.


\section{\label{sec:timeseries}Times series}

\begin{figure*}[!hbtp]
\centering  
\includegraphics[width=0.99\linewidth]{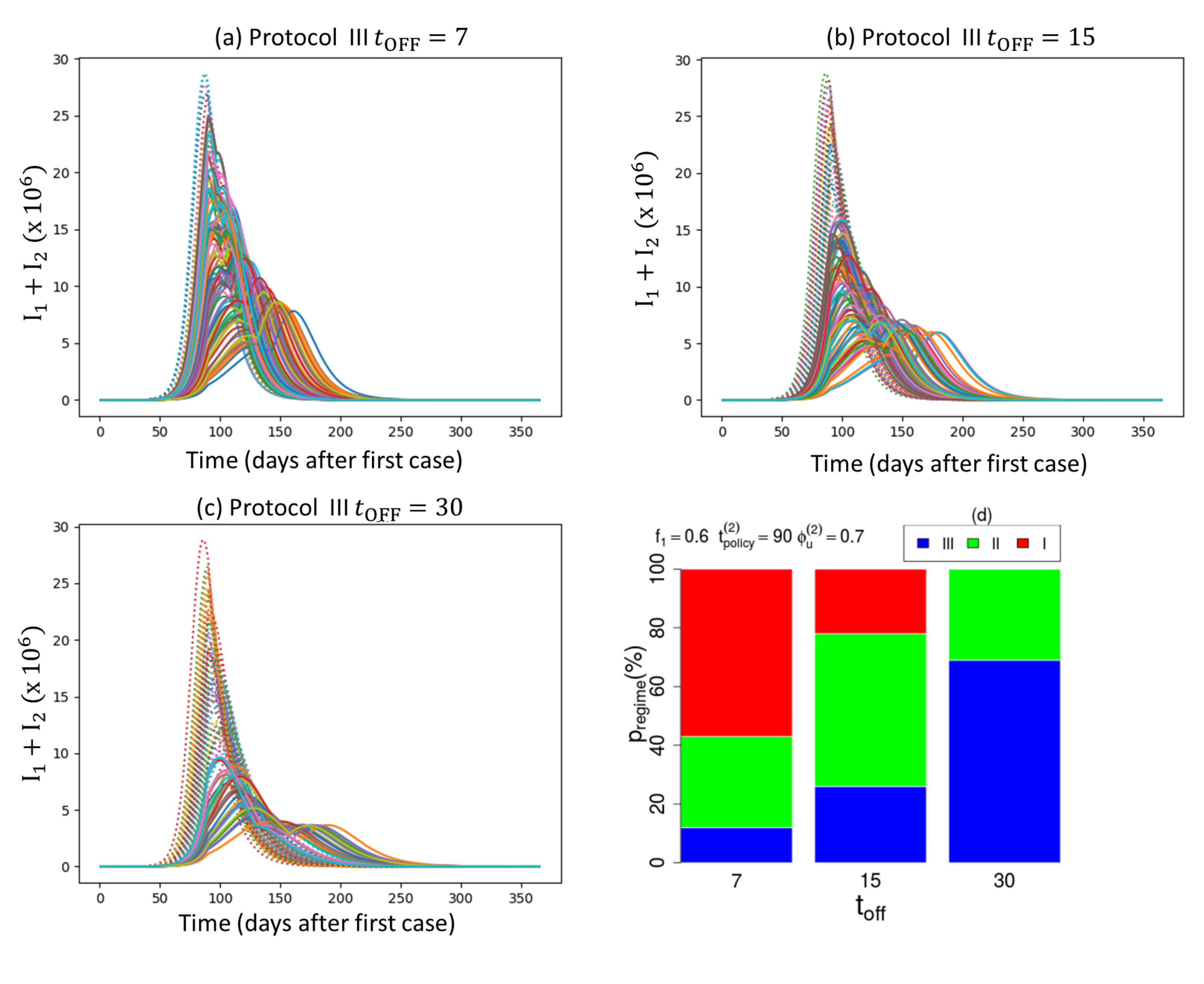}
\caption{ (a-c) Time series for the number of individuals in the class  $\sum_i I_i$ for $f_1=0.6$ as well as $\phi_1^{(3)}=0.8$ 
and $\phi_2^{(3)}=0.9$.}  
\label{Fig:time-series-br-1-matsup}


 
%
\includegraphics[width=0.99\linewidth]{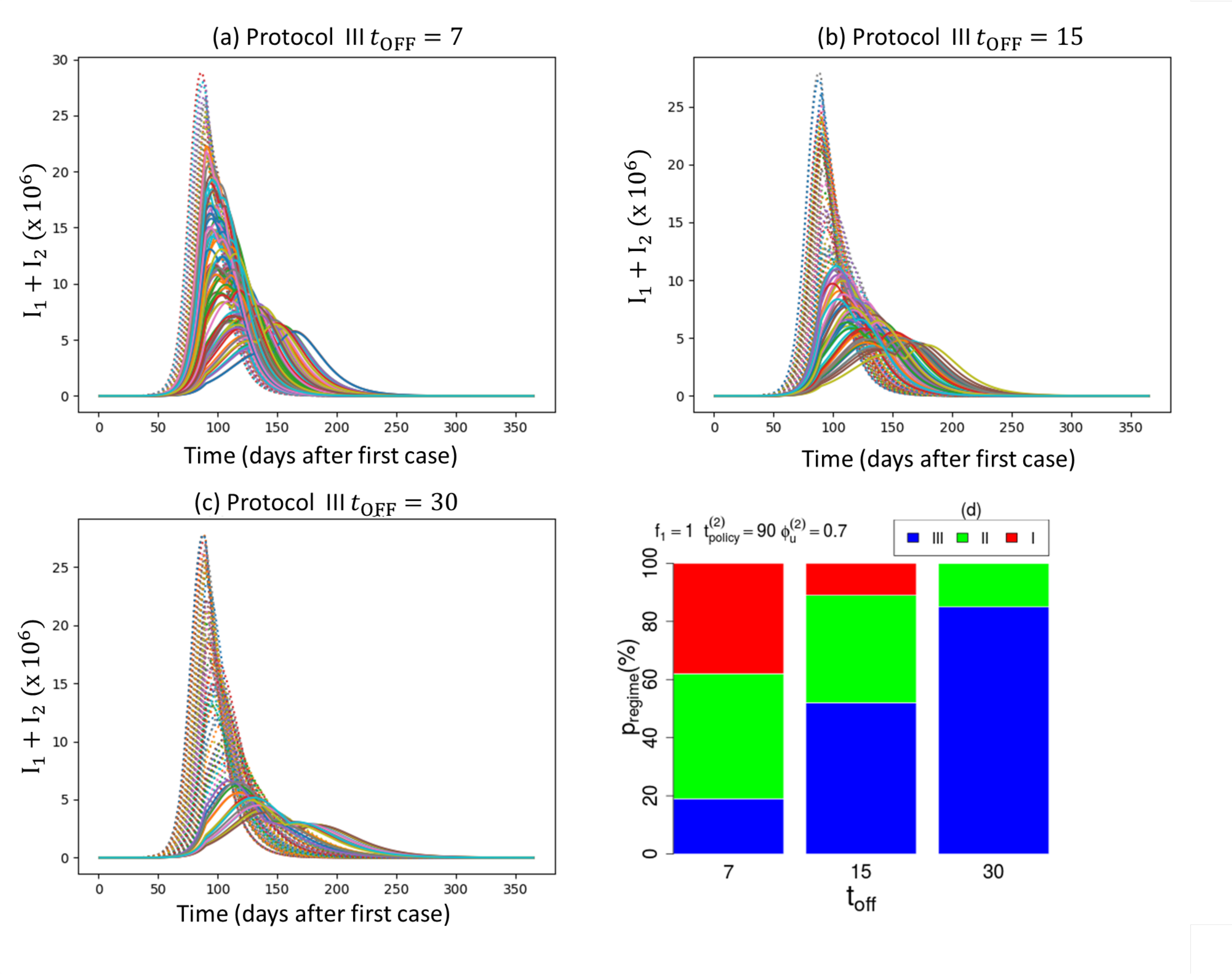}
\caption{ Same as Fig.\ref{Fig:time-series-br-1-matsup}, but for $f_1=1$.}  
\label{Fig:time-series-br-2-matsup}
\end{figure*}

 Figures~\ref{Fig:time-series-br-1-matsup}-\ref{Fig:time-series-br-2-matsup} 
 show the time series for number of infected agents.
 As stated in the main text, the noncompliance degree is decreased from 
 $\phi_u^{(1)}=0.799$ to  $\phi_u=\phi_u^{(2)}$ on day $t_{policy}^{(2)}=90$ after the first case at day $t_0$.
  Each panel shows 100 curves using parameters randomly sampled in the confidence interval provided previously\cite{bastos2020modeling}. 
 The self-isolation measures are lifted  $t_{OFF}$ days after the peak. At that moment the degree of noncompliance is increased to $\phi_1^{(3)}=0.8$ 
and $\phi_2^{(3)}=0.9$. The corresponding barplots in panels (d)  provide evidence  that the main results in our main manuscript are statistically robust, despite some noticeable fluctuations.

\begin{figure*}[!hbtp]
\centering  
\includegraphics[width=0.9\linewidth]{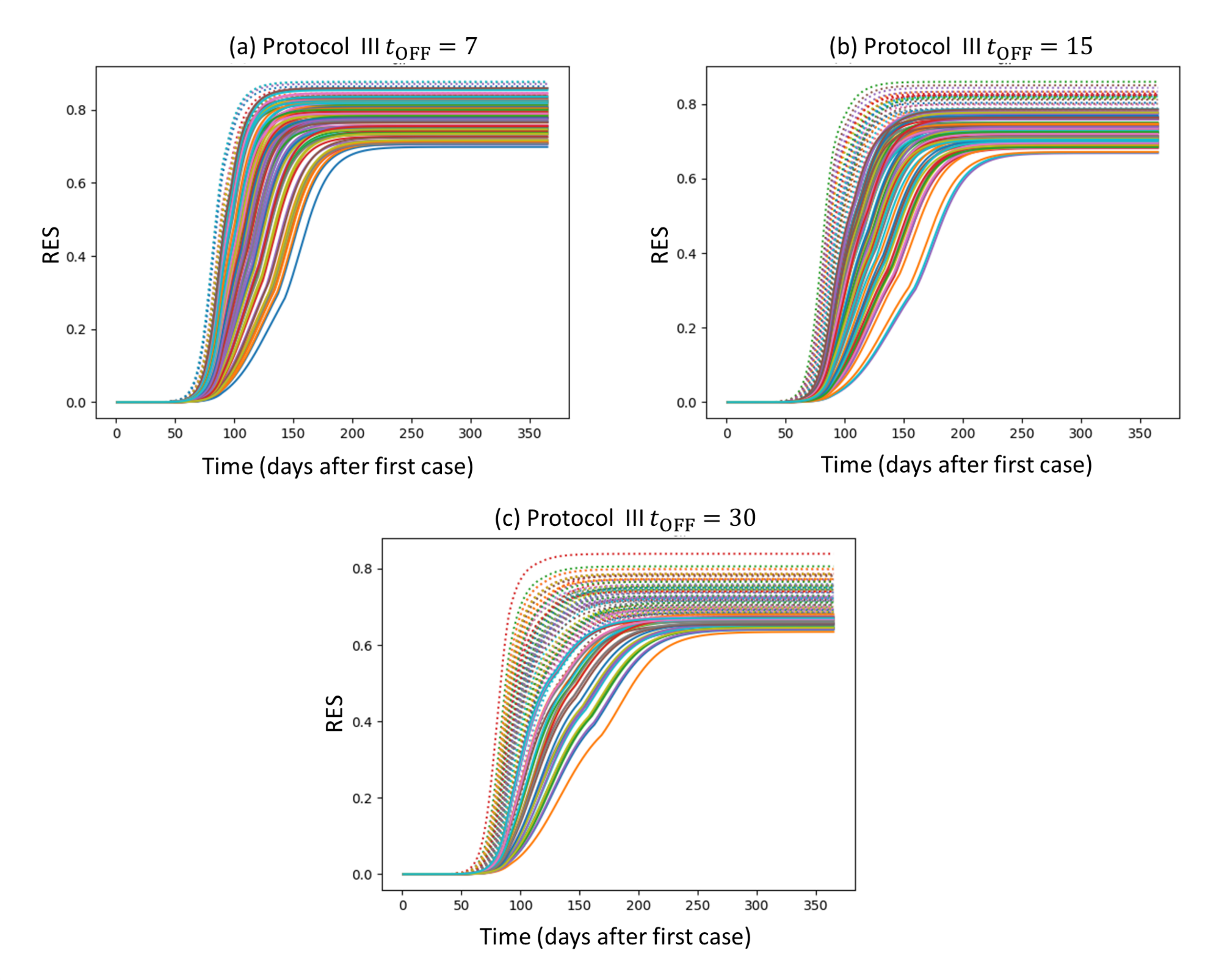}
\caption{Time series for  the Relative Epidemic Size (RES) corresponding to the outcomes presented in Fig.\ref{Fig:time-series-br-1-matsup}.}
\label{Fig:time-series-br-res-1}
\centering 
\includegraphics[width=0.8\linewidth]{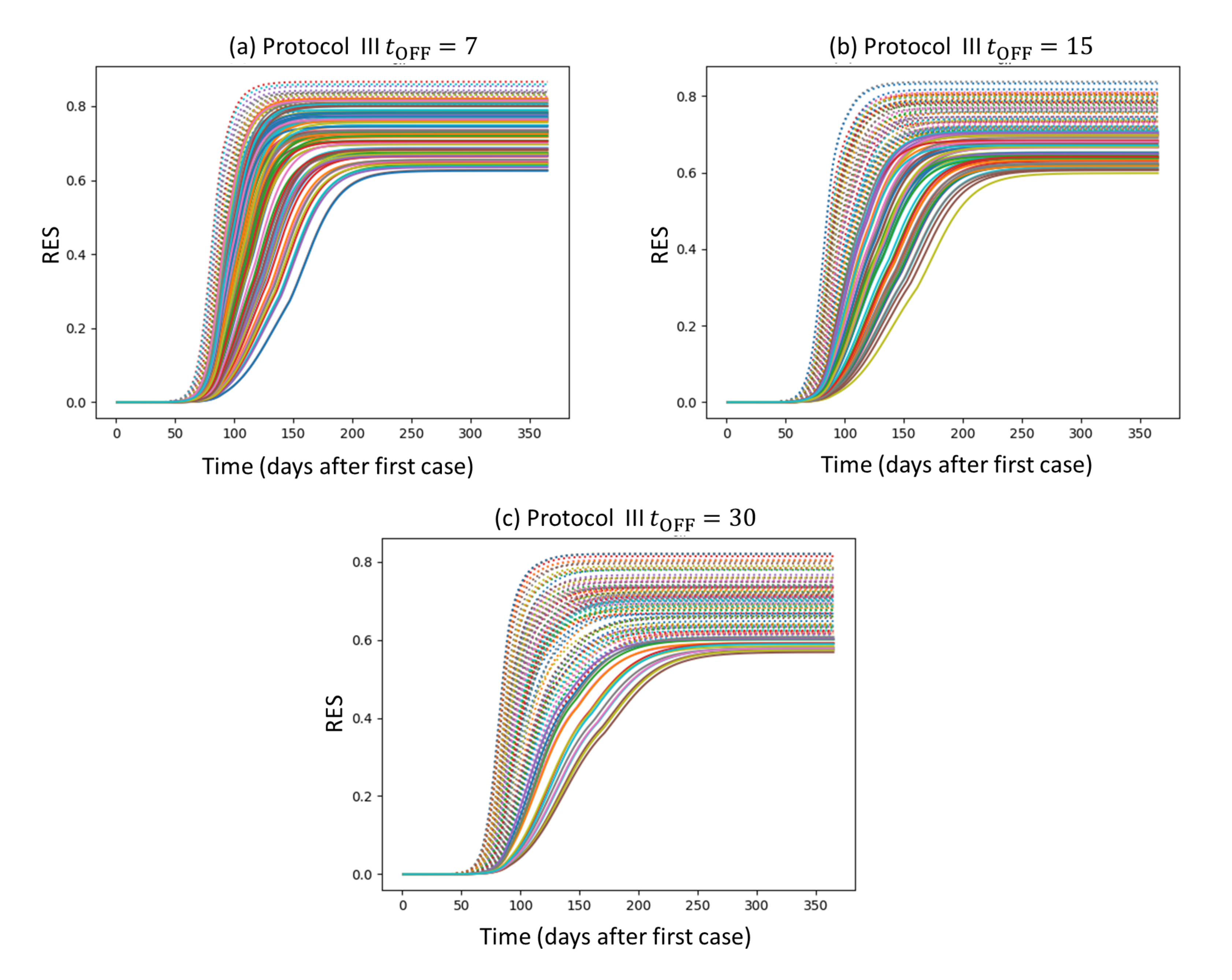}
\caption{Time series for  the Relative Epidemic Size (RES) corresponding to the outcomes presented in  Fig.\ref{Fig:time-series-br-2-matsup}. }
\label{Fig:time-series-br-res-2}
\end{figure*}

Figures~\ref{Fig:time-series-br-res-1}-\ref{Fig:time-series-br-res-2} show the time evolution for the
Relative Epidemic Size (RES)  computed from $t_0$ to $t$
\begin{equation}
RES \equiv \sum_{z=1}^{2}
\frac{S_z(t_0) - S_z(t)}{N}
\end{equation}
In those figures we see the standard behavior of the RES, i.e., the quantity grows with time as the susceptible population decreases by the raise of the number of infections. Most of those RES results can be linked to classical results which point to the quell of an infectious disease by herd immunity when the the level of (naturally) immunized people is between $60\%$ and $80\%$\cite{anderson1992infectious}. Nonetheless, we verify that some curves show saturation values below the former value. This finding of ours concurs with the results presented previously\cite{britton2020} where a heterogeneous population was considered as well.

\begin{figure*}[!hbtp]
\centering  
\includegraphics[width=0.7\linewidth]{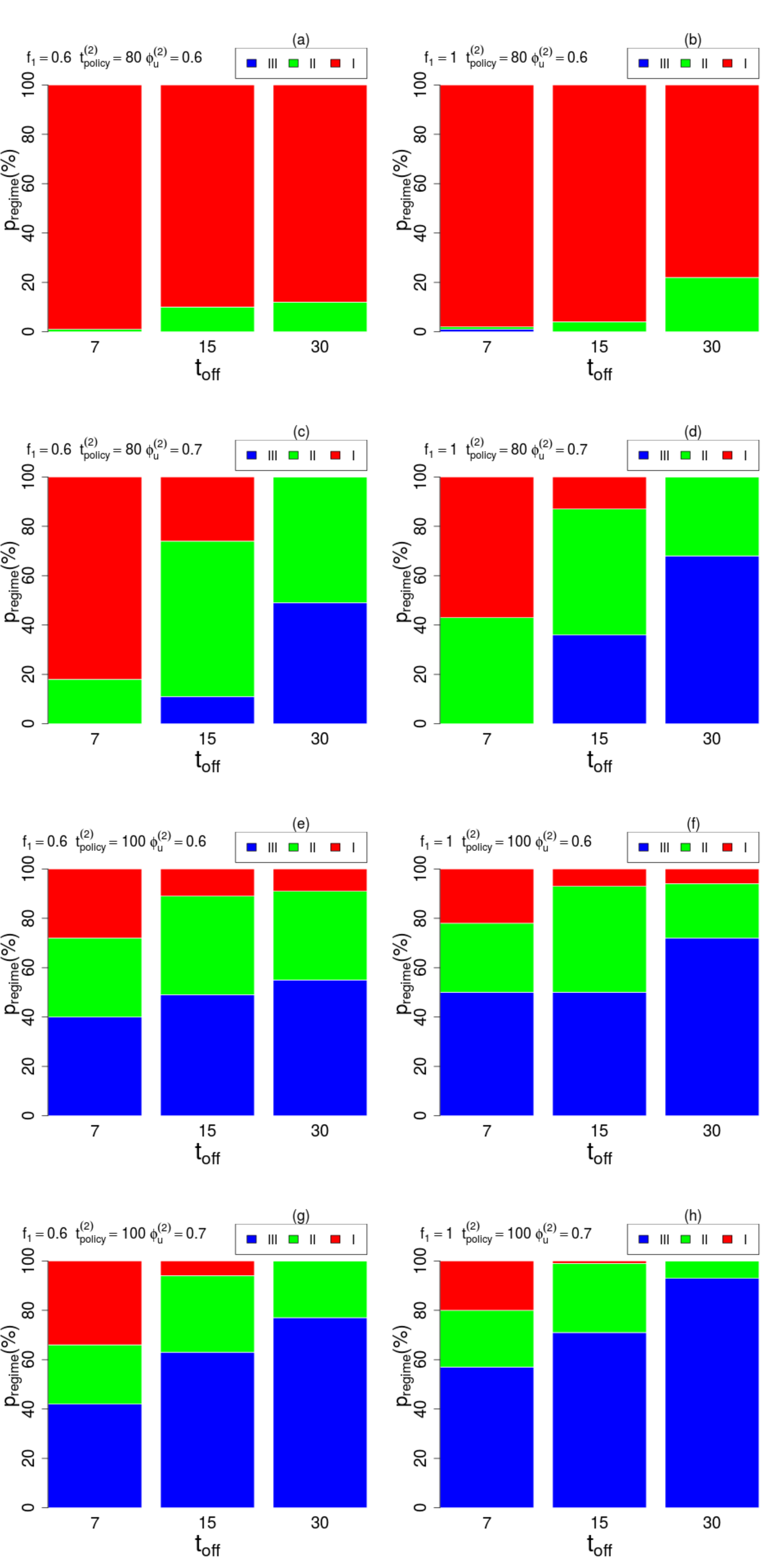}
\caption{Barplot for $t_{policy}^{(2)}=\{80,100\}$,  $f_1=\{0.6,1\}$ and  $\phi_u^{(2)}=\{0.6,0.7\}$. Other parameters: $\phi_1^{(3)}=0.8$ 
and $\phi_2^{(3)}=0.9$.  Outcomes for $100\times 3\times 8$ random samples.}

\label{Fig:barplot-4x4-1}
\end{figure*}

\begin{figure*}[!hbtp]
\centering  
\includegraphics[width=0.99\linewidth]{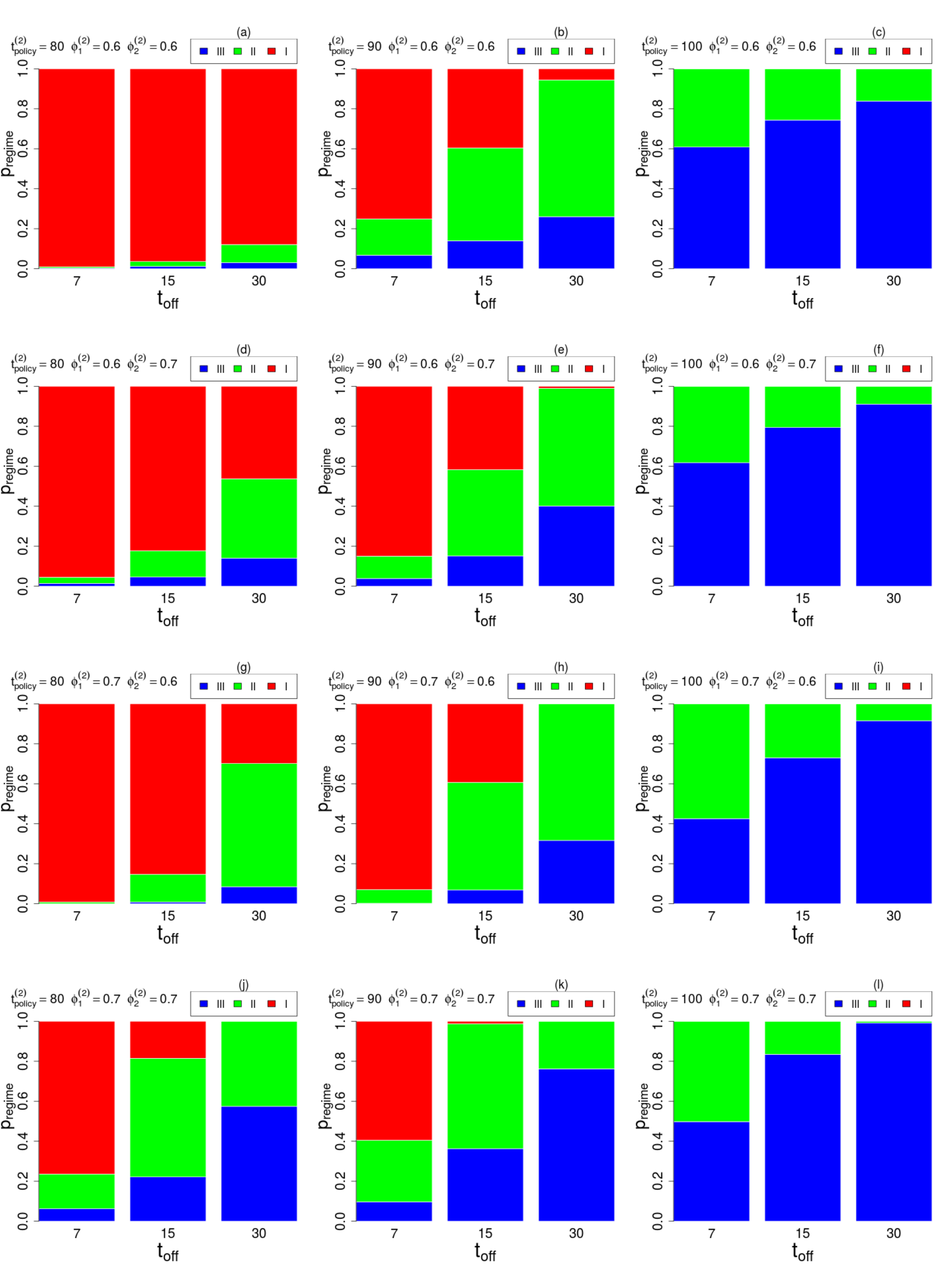}
\caption{Barplot with the proportion  of each regime $p_{regime}$ considering nonrandom combinations of $\phi_1^{(3)} \times \phi_2^{(3)} \in [\phi_1^{(2)},1] \times [\phi_2^{(2)},1]$ that satisfy $\phi_1^{(3)} \leq \phi_2^{(3)}$. We set  increments of $0.01$ for $\phi_u^{(3)}$.  Other parameters: $f_1=0.6$, $t_{policy}^{(2)}=\{80,90,100\}$.}

\label{Fig:barplot-2x2-phi1_2}
\end{figure*}

One can also discuss about variations in the noncompliance parameters $\phi_1^{(2)}$ and $\phi_2^{(2)}$, related to the time window $[t_{policy}^{(2)},t_{peak}+t_{OFF}]$, where $t_{policy}^{(2)}$ is the time from the first day where the noncompliance degree is decreased from $\phi_u^{(1)}$ to $\phi_u^{(2)}$ ($u=\{1,2\}$). In turn, $t_{peak}$ is the time to reach the peak. In Figure \ref{Fig:barplot-4x4-1}  we exhibit barplot graphics considering $t_{policy}^{(2)}=\{80,100\}$ as well as $f_1=\{0.6,1.0\}$ for fixed $\phi_1^{(3)}=0.8$ and $\phi_2^{(3)}=0.9$. In addition, in Figure \ref{Fig:barplot-2x2-phi1_2} we exhibit barplot graphics for $t_{policy}^{(2)}=\{80,90,100\}$, fixed $f_1=0.6$ and several combinations of the parameters $\phi_1^{(3)}$ and $\phi_2^{(3)}$ satisfying $\phi_1^{(3)}\leq \phi_2^{(3)}$. In this case, we are considering distinct number of days where the population compliance with isolation policies increases, as well as distinct values of such compliance. For the smaller value $t_{policy}^{(2)}=80$, the population spend less time with higher noncompliance degree $\phi_1^{(1)}=\phi_2^{(1)}=0.799$, and after the noncompliance decreases to $\phi_1^{(2)}$ and $\phi_2^{(2)}$ (the proportion of the population in isolation increases). In this case, we expect that a smaller number of individuals infected in comparison with the values considered in the main text. In such case, when the isolation policies are relaxed, we have more people "available" to be infected, and the second peak is most probable to be higher (regime I), even for $t_{OFF}=30$.

\vspace{0.5cm}

\begin{figure*}[!hbtp]
\centering 
\includegraphics[width=0.9\linewidth]{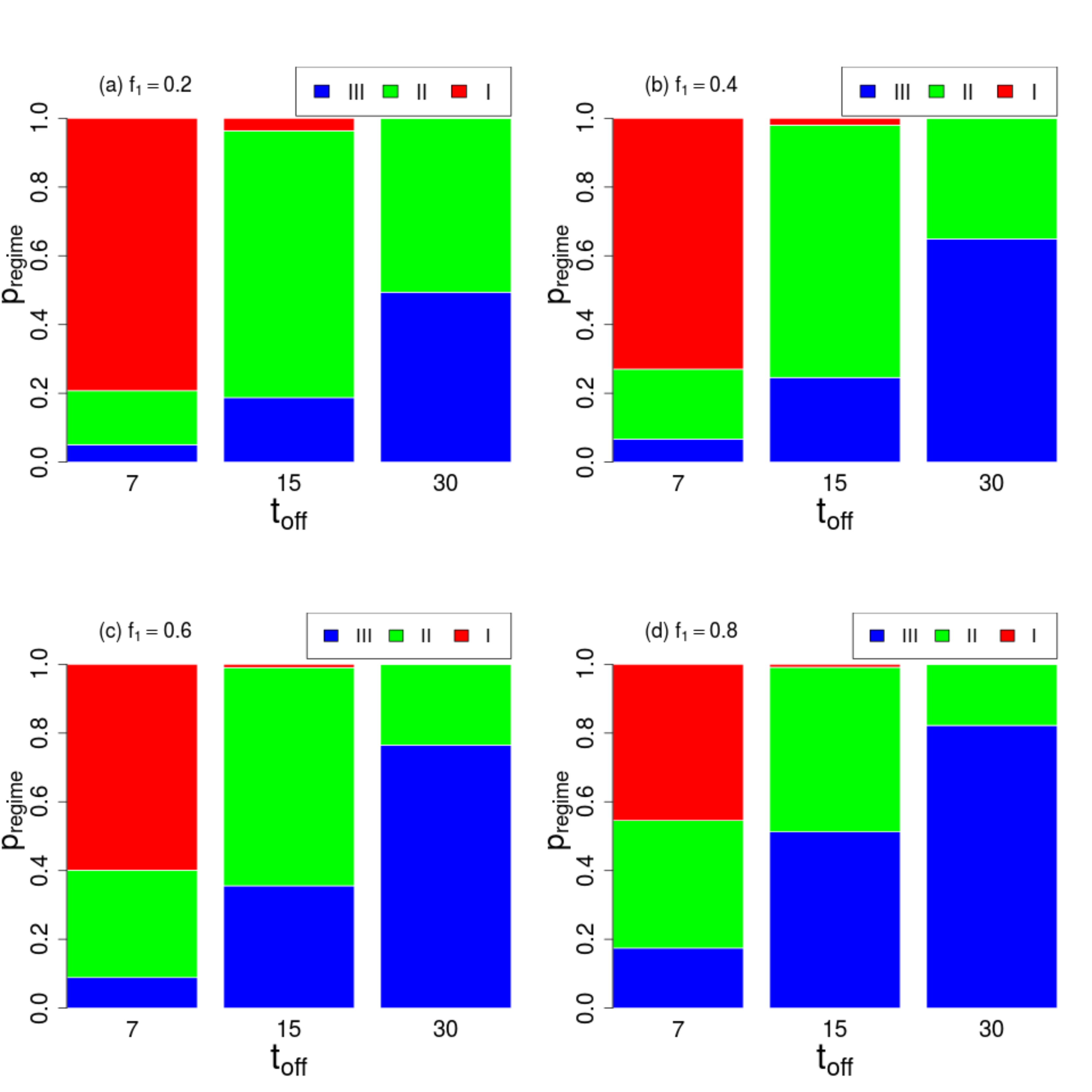}
\caption{Barplot with the proportion  of each regime $p_{regime}$ considering nonrandom combinations of $\phi_1^{(3)} \times \phi_2^{(3)} \in [\phi_1^{(2)},1] \times [\phi_2^{(2)},1]$ that satisfy $\phi_1^{(3)} \leq \phi_2^{(3)}$ as well as $f_1<f2$ and  $f_1>f2$. We set increments of $0.01$ for $\phi_u^{(3)}$. Other parameters: $t_{policy}^{(2)}=90$ and $\phi_u^{(2)}=0.7$.}

\label{Fig:barplot-2x2-2}
\end{figure*}

For further investigations about the model's parameters, we depicted in Figure \ref{Fig:barplot-2x2-2} barplot graphics considering the cases $f_1>f_2$ and $f_1<f_2$. One can see that the cases with small values of $f_1$ lead to a higher possibility of the occurrence of a larger second peak (region I, red).

\FloatBarrier

\providecommand{\noopsort}[1]{}\providecommand{\singleletter}[1]{#1}%

\end{document}